\documentstyle[aps,epsf]{revtex}

\draft
\begin{document}
\twocolumn[\hsize\textwidth\columnwidth\hsize
     \csname @twocolumnfalse\endcsname

\title{A realist view of the electron: recent advances
and unsolved problems}

\author{W. A. Hofer} 
\address{CMS - Vienna, Getreidemarkt 9/158,
A--1060 Vienna, Austria\\
{\bf email:} whofer@cms.tuwien.ac.at \hspace{0.8cm} {\bf web:}
info.tuwien.ac.at/cms/wh/}

\maketitle

\begin{abstract}
In view of experimentally obtainable resolutions, equal to
the Compton wavelength of an electron, the conventional interpretation of
quantum mechanics no longer seems to provide a sufficiently subtle tool.
Based on the intrinsic properties of extended particles we propose a new
theory, which allows to describe fundamental processes with unlimited
precision at the microlevel. It is shown how this framework combines
classical electrodynamics and quantum mechanics in a single and consistent
picture. An analysis of single measurement problems reveals
that the theory is suitable to remove some of
the most striking paradoxes in quantum mechanics, which are
found to originate from obscuring statistical effects with
physical reasoning. A possible origin of the infinity problems
in relativistic quantum fields is found by analyzing electron
accelerations due to photon
absorption processes. The current state of the theory
and existing problems are discussed briefly.
\end{abstract}

\pacs{03.65.Bz, 14.60.Cd, 11.10.Gh, 41.20.Gz}

\vskip2pc]

\section{Introduction}

As long as the size of elementary particles remained somewhat 
insignificant, the theoretical efforts could be limited 
to a description of point-like entities
with qualities like mass, charge, or spin. The theoretical
framework gradually evolving in quantum electrodynamics (QED)
or quantum field theory \cite{schweber94,schweber61} consequently 
knew little more than these point-particles and their interactions.
But experimental methods have developed rapidly. On the one hand, the 
precision of measurements reaches already far
into a regime comparable to the size of particles. The
Compton wavelength of an electron, for example, is not smaller
than current resolutions in surface science.
On the other hand, quantum effects play a major role
also in experiments, where the length scale is in the range
of centimeters or even meters. This situation makes the 
development of consistent models of particles, taking into
account the relationship between classical and quantum systems,
an even more important issue.

Following the traditional method of development, one starts with
the definition of e. g. an electron, while the relation between
electron properties and electromagnetic fields is determined in
a second step. On this basis the treatment of the problem could
be commenced with Barut's assessment of the main problem in
electron theory \cite{barut97}:
{\em If a spinning particle is not quite a point-particle, 
nor a solid three dimensional top, what can it be?}  
According to Bunge or Recami \cite{bunge55,recami97}
there are in general only three possibilities: 
A particle either is (i)
strictly point-like, (ii) actual extended, or (iii)
a point-like structure in motion within the actual volume
of the particle. And the only possible solution for an electron
is of type (iii).

The thesis elaborated in this paper uses a different method,
which can be described by four assertions: (i) An electron is
determined by its intrinsic properties. (ii) The exact numerical
value of these properties, and consequently the actual size of
the electron remains undefined. (iii) The statistical ensembles in
quantum theory (QT) are based on this indefiniteness. (iv) 
Quantized properties like mass, charge, or spin are due to a
change of intrinsic properties in the presence of external fields.
Although the picture is far from complete, it suggests a 
modification of current concepts in the following sense: even
if there are intrinsic properties of single entities called
electrons, there may not be single and well defined {\em objects}
called electrons. 

Since the ensembles in QT are related to electrons with a defined
range of intrinsic properties, the interaction of a member of the
ensemble with exactly determined intrinsic properties is 
exactly determined. We use the term {\em elementary process}
in such a case, and the statistical results of measurements in
QT are thought to contain an arbitrary number of such
processes. This definition of elementary processes
implicitly contains the assumption of hidden variables,
although the present approach is substantially different
from Bohm's \cite{bohm52} theory: (i) The 
notion of a particle is not fundamental. 
(ii) We start with a description of intrinsic properties
and relate them only later to the formulations in QT 
and classical ED.
(iii) In this way we regain a statistical
(and non-local) interpretation of QT, where the statistical ensembles
result from an unknown phase (similar to Bohm's picture),
but also, due to the uncertainty relations, from an unknown
energy component. 

Especially the latter quality of the ensembles in QT accounts 
for a fair share of the paradoxes, which have been irritating 
- or exciting - the physical community for quite some time. And
it is exactly this quality of the quantum ensembles which makes
QT an {\em incomplete} theory. The results
are described in view of a non-specialized readership, 
specialist readers are referred to existing publications, 
describing all the necessary steps in great detail 
\cite{hofer95,hofer97a,hofer97b,hofer97c,hofer97d,hofer98a,hofer98b,hofer98c}.

\section{The electron: intrinsic properties}

There can be no doubt that the electron, of all the
elementary particles, is by far the best researched,
both experimentally and theoretically. Should it therefore
be a mere exaggeration, if a book by MacGregor,
appearing in 1992, is entitled: The Enigmatic Electron
\cite{mcgregor92}. Or is there substance to this claim? 
The main problem with all existing models of the electron
(see the introduction) is that none of them can explain all
the observed experimental features. An extended particle,
for example, is claimed to contradict scattering experiments,
a point-particle, on the other hand, leads to the well known
infinity problems in QED. Both of these problems have, in
the model about to be described, a common solution, which
can only be systematically displayed, if we focus our 
attention on the {\em intrinsic properties} of an electron,
while we shall recover the solutions to the above problems
only afterwards. 

We start with a non-relativistic frame of reference, 
assuming that the electron has a finite volume $V$, which 
is not specified, and that it moves with constant
velocity $\vec{u} = u \vec{e}_{u}$. 
Its intrinsic structure is described by
a wave equation for its density of mass $\rho (\vec{r},t)$, 
and a wave equation
for an additional field energy $\phi (\vec{r},t)$, 
which was called the
{\em intrinsic potential} \cite{hofer98b}
for two reasons: First, the same model can be
used for photons (in this case the particles proceed
with $c$), the intrinsic field energy $\phi (\vec{r},t)$ then
is equal in magnitude to the electromagnetic potential
$\vec{A}$. And second, due to electron propagation the energy is 
shifted from the propagating mass  
to the correlating intrinsic field energy
and vice versa, the field energy therefore behaves 
like a periodic potential.

The impact of this concept on the fundamental statements
in QT is quite substantial: (i) The energy of the electron
is double the kinetic energy or equal to $m_{e}u^2$.
(The energy is computed by integrating
intrinsic properties over the volume $V$. The exact value of
$V$ is not required for the integration.)

\begin{eqnarray}
W_{kin} & = & \frac{1}{2} m_{e} u^2 \qquad
W_{pot} = \frac{1}{2} m_{e} u^2 \nonumber \\
W_{tot} & = & W_{kin} + W_{pot} = m_{e} u^2 = \hbar \omega
\end{eqnarray}

If the total energy is used to satisfy Planck's relation,
the dispersion relations for monochromatic plane waves
are recovered, then (ii) regarding its intrinsic properties,
an electron can be described as a monochromatic plane wave.
But this means that (iii) the Schr\"odinger equation \cite{schrodinger26}, 
which neglects the intrinsic energy components,
is no longer an exact equation \cite{hofer98b}.
And on this basis it can be deduced that (iv) the
Heisenberg uncertainty relations \cite{heisenberg27}
actually describe the errors due to the omission of
intrinsic energy. 

The latter point is interesting for three separate reasons.
First, it is well known that the uncertainty relations are responsible
for the spreading of a wave packet (see e.g. 
\cite{debroglie82}). If they are not interpreted as
physical causes - as they are thought to be in the standard 
framework, even if the more innocent term {\em principle} is used - then
the spreading of a wave packet is not a physical effect, but
only a consequence of the logical structure of QT. Second, the
result seems to settle the long-standing controversy
between the {\em axiomatic} and the {\em empirical} 
interpretation of Heisenberg's relation. It is not
empirical, since it does not depend on any measurement
process; but it is also not a physical principle, because it
is due to the fundamental assumptions of QT. 
Third, if the uncertainty relations are not a universal
physical principle, then experiments can 
be described with unlimited precision also
at the microlevel: only in this case does the definition of
{\em fundamental processes} at all make sense.

\begin{figure}
\begin{center}
\begin{tabular}{r}
\epsfxsize=0.9\hsize
\epsfbox{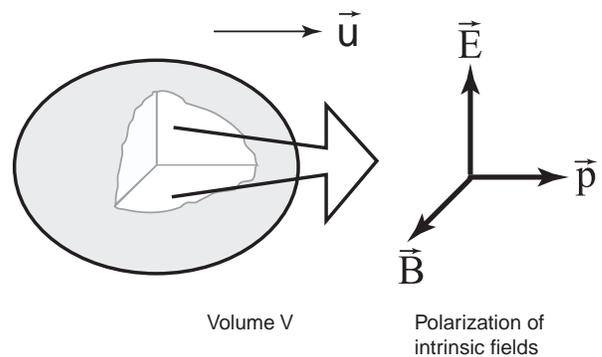}
\end{tabular}
\end{center}
\vspace{0.5cm}
\caption{Intrinsic properties of electrons.
The momentum density $\vec{p} = \rho \vec{u}$ is longitudinal, while
the electromagnetic $\vec{E}$ and $\vec{B}$ fields are of
transversal orientation. The model is valid for electrons and
photons, and it can be deduced that the intrinsic magnetic field
$\vec{B}$ is responsible for the magnetic moment of electrons.}
\label{intr_fig}
\end{figure}

So far the proposition of intrinsic structures and intrinsic
properties is a mere speculation, which is not very different
from other speculations including a more or less substantial
part of the known qualities of electrons into a single picture.
In particular it is yet unclear, how magnetic properties come
into play. This is done in two steps: (i) The intrinsic potential
$\phi$ is interpreted as an {\em electromagnetic} property, 
related to intrinsic
electromagnetic fields $\vec{E}$ and $\vec{B}$ of transversal 
orientation, and which comply with a wave equation.
The direction of intrinsic fields is shown in
Fig. \ref{intr_fig}. 
And (ii) it is proved that these conditions are generally
sufficient to derive Maxwell's
equations \cite{jackson84}. Therefore, the intrinsic structures
lie at the bottom of two hitherto separate concepts:

\begin{itemize}
\item They are the origin of wave-like qualities of the electrons
described by Schr\"odinger's equation.
\item And they are the origin of electromagnetic qualities of
electrons (and photons), since they lead to our - to date -
best theory of electromagnetic fields.
\end{itemize}

The peculiar features of spin in QT can only be fully understood from
interactions of electrons with external fields. This will be done in
the following sections. Here we wish to add a few
remarks on the measurements of Bell type inequalities, performed
in ever increasing perfection since the Eighties
\cite{aspect82}, and which seem
to indicate what is usually called {\em action at a distance}.

It is a common error, especially among experimenters,
to assume that these measurements demonstrate that
{\em nature} allows action at a distance, that nature is
{\em non-local}. Without claiming that this is impossible, it can 
nevertheless be said that it cannot be proven by these
- EPR type - measurements due to the fact that a valid
measurement of spin correlations violates the uncertainty
relations. Why is this decisive? In order to understand the
argument, let us analyze the theoretical basis of
these measurements. Interpreting a measurement of spin, spin has
to be defined, and it can only be defined within QT, but not
in classical electrodynamics. For EPR like experiments 
we also require a conservation principle, since the
total spin of two particles must be known. It could be
analyzed, furthermore, what theoretical basis is required 
for the deduction of the Bell inequalities, and it can be
argued, that some of the assumptions going into these
deductions are a lot less important than locality in physics
(see e.g. \cite{claverie80,kracklauer98}). This is not needed
in the present context, since it can be shown that no valid
interpretation of spin correlation measurements within QT is
possible. To this aim we define the spin of a particle by
the magnetic moment $\vec{\mu}$ and a magnetic field $\vec{B}$,
which shall be determined from intrinsic properties
($\vec{u}$ is the velocity, $\rho$ the density of the electron
as previously defined):

\begin{eqnarray}
W & = & - \vec{\mu} \vec{B} = \frac{1}{2} \hbar \omega 
\qquad \vec{\mu} = g \frac{e}{2 m} \vec{s} \nonumber \\
\vec{B} & = & - \frac{1}{2 \cdot \bar{\sigma}} \nabla \times \rho \vec{u}
\end{eqnarray}

$\bar{\sigma}$ in these relations is a dimensional constant to make 
mechanical units compatible with electromagnetic variables.
Assuming that the kinetic energy $\frac{1}{2} \hbar \omega$ 
is due to the interaction of a constant magnetic moment
with the intrinsic magnetic field of the electron,
we arrive at  $g_{e} = 2$ and  $s_{e} = \hbar/2$,
while the direction of spin is equal to the direction of
the intrinsic magnetic field \cite{hofer98b}. For photons
the same calculation yields $g_{ph} = 1$ and $s_{ph} = \hbar$.
These results will be clarified in the calculation of
interactions with external fields further down.

If we now consider a correlation
measurement of photon spin we are confronted with the problem
that spin is parallel to the intrinsic magnetic field, which is
a periodic variable: spin therefore cannot be constant but will
oscillate from $+ s$ to $- s$ with a period of half the particle's
wavelength. For a valid correlation measurement the local precision
therefore must be higher than $\lambda/2$. But it has been 
demonstrated, in the deduction of the uncertainty relations
via the omission of intrinsic potentials \cite{hofer98b}, that
this is the highest limit of precision possible in QT; thus a
valid measurement {\em exceeds} the level of precision 
provided for in QT, thus it is incompatible with the axioms
of QT: it can therefore not be consistently interpreted within
this same framework. Independently of any other consideration.
And since EPR measurements rely on QT for the definition and
conservation of spin, they are generally inconclusive.

Returning to Barut's dilemma quoted in the introduction, it can
be said that in this model the electron is neither a spinning top nor any 
modified point particle: it is an extended 
structure, and so far it is not clear, which of the
intrinsic properties  of the electron
is actually related to its charge.

\section{Interactions with external fields}

To elucidate the problem, let us consider the interaction of
an electron (density amplitude $\rho_{0}$, charge density
amplitude $\sigma_{0}$) with a photon 
(density amplitude $\rho_{ph}$) under the presence of an external 
electrostatic field $\phi_{ext}$. The procedure used for the calculation is
pretty standard: we define the Lagrange density of an electron
in motion, including an external potential and a presumed photon.

\begin{equation}
{\cal L} := T - V = \rho_{0} \dot{x}_{i}^2 + 
\rho_{ph} c^2 - \sigma_{0} \phi_{ext}
\end{equation}

A variation with fixed endpoints and using the
principle of least action allows to calculate, by way of
a Legendre transformation and in a first order approximation of
a Taylor series, the Hamiltonian of the system as
\cite{hofer98b}:

\begin{equation}
H = \frac{\partial {\cal L}}{\partial \dot{x}_{i}} \dot{x}_{i} -
{\cal L} \approx \sigma_{0} \phi_{ext}
\end{equation}

The result seems paradoxical in view of kinetic energy of the
moving electron, which does not enter into the Hamiltonian.
Assuming, that an electron is accelerated in an
external field, its energy density after interaction with
this field would only be altered according to the change of
location. The contradiction with the energy principle is
only superficial, though. Since the electron will have been
accelerated, its energy density {\em must} be changed. If
this change does not affect its Hamiltonian, the only
possible conclusion is, that photon energy has equally
changed, and that the energy acquired by acceleration
has simultaneously been emitted by photon emission.
Therefore, the initial system was over--determined, and
the simultaneous existence of an external field {\em and}
interaction photons is no physical solution to the
interaction problem. It should be noted that the conclusion
is only valid in the first order expansion used, an approach
which was necessary due to the unknown relation between photon density
and electron velocity.

In this case the process of electron acceleration must be
interpreted as a process of simultaneous
photon emission: the acquired kinetic energy is balanced
by photon radiation.
A different way to describe the same result
would be saying that electrostatic interactions are
accomplished by an exchange of photons: the potential of
electrostatic fields then is not so much a function of
location than a history of interactions.
This can be shown by calculating the interaction Hamiltonian
$H_{w}$:

\begin{eqnarray}
H_{0} = \rho_{0}\, \dot{x}_{i}^2 + \sigma_{0} \phi
\qquad
H = \sigma_{0}\, \phi \nonumber\\
H_{w} = H - H_{0} = - \rho_{0} \,\dot{x}_{i}^2 =
\rho_{ph} \,c^2
\label{ep025}
\end{eqnarray}

But if electrostatic interactions can be referred to an exchange of
photons, and if these interactions apply to accelerated electrons,
then an electron in constant motion does not possess an intrinsic
energy component due to its electric charge: electrons in constant
motion are therefore stable structures.

The photon-interaction model of electrostatic fields allows a
further extension of the current framework. Since, what might
be called {\em charge}, finds its expression in the properties
of the emitted and absorbed photons, and since these photons
can either lead - by way of their intrinsic properties - to
attraction or repulsion of other charges, the sign of the charge
is no longer a quantity fixed for all time and under every 
condition. Although there is, currently, no comprehensive way
of describing the {\em origin} of a specific charge/anticharge
in a specific situation, it seems that the model should also
be suitable for questions of this type and which are well beyond
the rather phenomenological (Heisenberg) description used in
the current standard models of elementary particles. 

As a second example we calculate the interaction of an electron 
with an external magnetic field. The field in this case changes
the intrinsic energy components of the electron.
The {\em local} and {\em deterministic} calculation of these
interactions is based on the field equations of intrinsic fields.
The units of these fields are due to the derivation
of the Maxwell equations from intrinsic properties, an analysis of
electromagnetic units has been given in \cite{hofer96}:

\begin{eqnarray}
\frac{1}{u^2} \frac{\partial \vec{E}}{\partial t} & = & \nabla \times
\vec{B} \qquad 
- \frac{\partial \vec{B}}{\partial t} = \nabla \times \vec{E} 
\nonumber \\
\phi_{em} & = & \frac{1}{2} \left(
\frac{1}{u^2} \vec{E}^2 + \vec{B}^2 \right)
\end{eqnarray}

We use a dynamic model by assuming that the external magnetic
field is switched on in a finite interval $[0,\tau]$. Then the
intrinsic energy component in the magnetic field has changed
and will be \cite{hofer97c}:

\begin{equation}
\phi (\vec{B}_{ext}) = \phi_{em} + |\vec{B}_{ext}|^2
\end{equation}

\begin{figure}
\begin{center}
\begin{tabular}{|c|}
\hline
\epsfxsize=0.9\hsize
\epsfbox{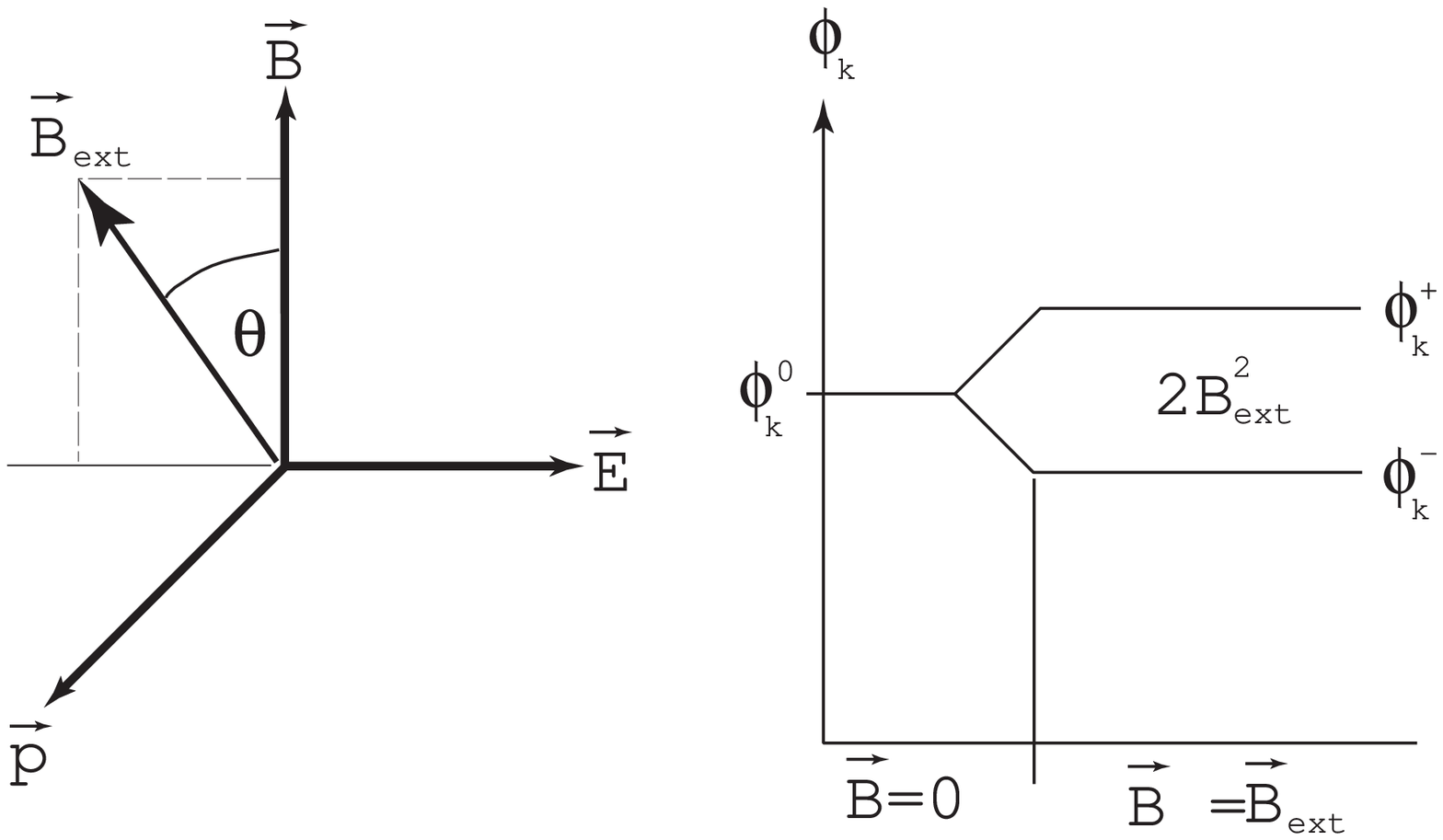} \\
\hline
\epsfxsize=0.9\hsize
\epsfbox{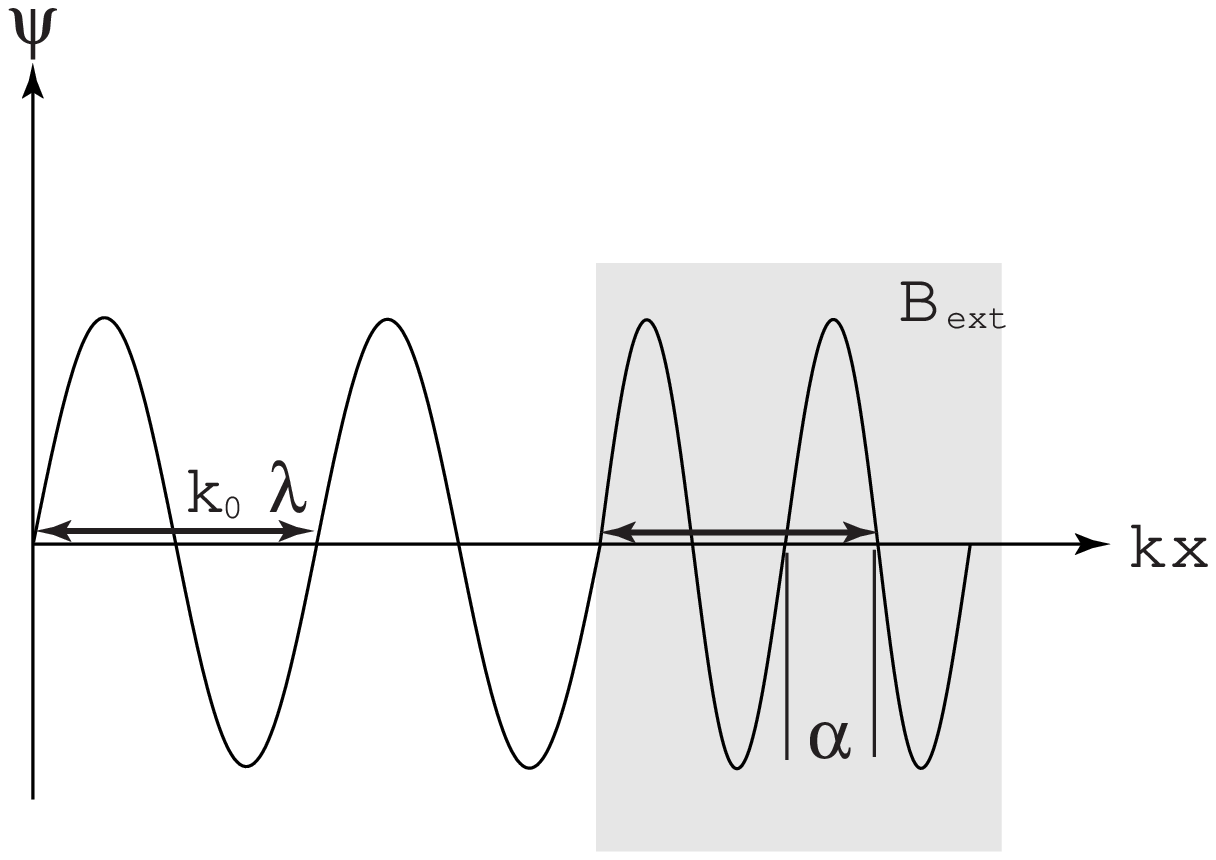} \\
\hline
\end{tabular}
\end{center}
\caption{The kinetic energy component $\phi_{k}$ is changed
in a magnetic field
due to interaction, the change is independent of the
angle $\theta$ between the external and the intrinsic magnetic
field components (top). Due to the interaction the wavelength of
the particle wavefunction is changed (bottom).}
\label{magn_intr}
\end{figure}

The crucial feature of magnetic interaction is, that the acquired
energy is independent of the angle $\vartheta$
between the intrinsic magnetic field $\vec{B}$ and the
external magnetic field $\vec{B}_{ext}$ (see Fig.\ref{magn_intr}).
It can therefore not be formalized as the scalar product of an
intrinsic (and constant) magnetic moment $\vec{\mu}$ and an
external field $\vec{B}_{ext}$:

\begin{equation}
W \ne - \vec{\mu} \cdot \vec{B}_{ext} \qquad 
\vec{\mu}, \vec{B}_{ext} \in R^3
\end{equation}

or only, if the magnetic moment is a non-local variable:
the non-local definition of particle spin in quantum theory,
or the impossibility to describe
spin as a vector in $R^3$, can only be understood on the
basis of interactions. More specifically, it is the 
- failed - attempt to describe the changes due to magnetic 
interactions with a formulation inherited from classical 
electrodynamics. Therefore, spin in quantum theory {\em cannot} be a vector,
{\em because} interactions do not depend on the direction
of field polarization. This result, which only applies to free 
particles, also illustrates the importance
of dynamic models of interactions.

As the intrinsic component of particle energy
is increased due to external fields, total
energy of the particle can either be increased, which should be
the case for charged particles like electrons, or it can remain
constant, which we, tentatively, assume for neutral particles 
like neutrons. In any case the kinetic
components of particle energy change, and this change corresponds
to a changed wavelength of its wavefunction $\psi$. If the change
of the wavelength is calculated and the phase difference to a
beam not subjected to this magnetic field estimated, we arrive
at the following  result for the phase-difference $\alpha$:

\begin{equation}
\alpha = 2 \pi \left(
\frac{l}{\lambda} \frac{|\vec{B}_{ext}|}{\sqrt{\bar{\rho} u^2}} - n
\right) \qquad n \in N
\end{equation}

where $l$ is the path-length in the field of the magnet and
$\lambda$ the original wavelength of the beam. The result
indicates that the phase difference is linear with the 
intensity of the magnetic field: a result confirmed by 
neutron interference measurements of Rauch and 
Zeilinger \cite{rauch92,zeilinger81}.

A short remark is in place concerning the relation 
of the present concept to the concept of {\em quantization}
in QT, which seems to be used in various, and
sometimes incompatible meanings. Provided, the structure of
matter consists of atoms, every detector and consequently
every measurement only yields discrete results: there must,
necessarily, exist a threshold which is required to trigger
a reaction. On the level of measurements quantization is
consequently a necessity (this also applies, for example,
to Millikan's experiments to determine the ''elementary 
charge''). But while this is trivial in the atomic domain,
it does not mean that we have to encounter discrete quantities
in a point-like volume when electrons are separated from atoms:
given the infinity problems connected with such an idea, it
seems amazing that it prevailed for so long, even in de Broglie's
''double-solution''. The only real argument, which suggests
such an approach, originates from scattering experiments:
if electrons were extended structures (three dimensional 
aggregations of mass) like atomic nuclei, then the scattering
cross section would be affected. This has never been observed
and it was concluded, therefore, that electrons cannot be 
extended structures (see e.g. Bender {\em et al.} \cite{bender84}). 
Comparing with the particle models 
introduced previously (see the introduction), 
these experiments only exclude electrons
of type (ii): three dimensional tops. They do not exclude any
other type, especially not the extended electrons
introduced in this paper, where interactions apply to every single
point of the internal structure.

It shall not be hidden, though, that a mathematical model for
scattering processes based on photon interactions has yet to be
developed: not an easy task, it seems, since the reduction to a
one body problem in a potential, like in standard solutions,
is not generally applicable.

\section{Ensembles in quantum theory}

It was already noted by David Bohm that QT does not differ
between elementary processes (or physical interactions) and
statistical results \cite{bohm66}:

{\em Yet it is not immediately clear how the ensembles,
to which ... probabilities [in QT] refer, are formed and
what their individual elements are. For the very terminology of 
quantum mechanics contains an unusual and significant
feature, in that what is called the physical state of an
individual quantum mechanical system is assumed to 
manifest itself only in an ensemble of systems.}

The Copenhagen interpretation seeks to make up for this
conceptual deficiency by asserting that {\em nature itself}
is the origin of this feature. However, we shall try in 
this section to determine the exact borderline between the
elementary processes and the statistical picture in QT.
As will be seen presently, the (unusual) statistics of
quantum systems have two separate origins: (i) The 
unknown intrinsic energy components. (ii) Normalization
of the wave function. The first accounts for the change of
the ensemble structure in measurements, since it affects
the range of allowed intrinsic energies, the latter 
introduces non-locality into the framework of QT, because
normalization requires an integral of the wavefunction
over the whole system considered: after normalization
the amplitude of the wavefunction in one region of the
system depends on the potentials and amplitudes of 
the wavefunction in all the other regions of the system.
QT therefore {\em cannot} be a local theory.
Which does not mean, as demonstrated above, that {\em nature}
itself must be non-local. 

Starting with the ensemble structure in QT pertaining to the
omission of intrinsic energy components, let us first consider
the situation of a free particle. In this case the external
potential $V(\vec{r})$ is zero, and the maximum of $k$, in
a plane wave basis of possible solutions of the Schr\"odinger
equation for fixed total energy $E_{T}$, is described by
$k^2 = m E_{T}/\hbar^2$. Since the phase of the wave-like
intrinsic components is unknown, the total energy can be
distributed in an unknown manner between the kinetic components,
described in QT, and the electromagnetic components, not 
considered in QT. At a specific point $\vec{r}$ of our system 
this means, that we are dealing with a Fourier integral over
an allowed range of states, which we called the {\em quantum
ensemble} of free electrons \cite{hofer97d}:

\begin{eqnarray} \label{eq010}
\psi(\vec{r}) & = &  \frac{1}{(2 \pi)^{3/2}} \int_{0}^{k_{0}}
d^3 k \psi_{0}(\vec{k}) \exp i \vec{k} \vec{r} \nonumber \\
k_{0} & = & \sqrt{\frac{m}{\hbar^2} E_{T} } \qquad
E_{T} = m u^2
\end{eqnarray}

where $\psi_{0}(\vec{k})$ is the $k$-dependent amplitude. 
This quantum ensemble, which is defined according to the
omission of intrinsic energy and thus according to the
uncertainty relations (see above), describes a range of allowed
kinetic energies, and it applies to every single point of
a given system. In this sense the {\em wavefunction} 
$\psi(\vec{r})$, given by Eq. (\ref{eq010}), is a statistical
measure \cite{born26}. The {\em unusual feature}, Bohm refers
to is thus, on closer scrutiny, removed: although the ensemble
is an integral part of QT, it does not mean, that we cannot 
go beyond the purely statistical picture of e.g. the
Copenhagen interpretation \cite{cramer86} to an analysis
of the underlying fundamental processes.

\begin{figure}
\begin{center}
\begin{tabular}{|c|}
\hline
\epsfxsize=0.9\hsize
\epsfbox{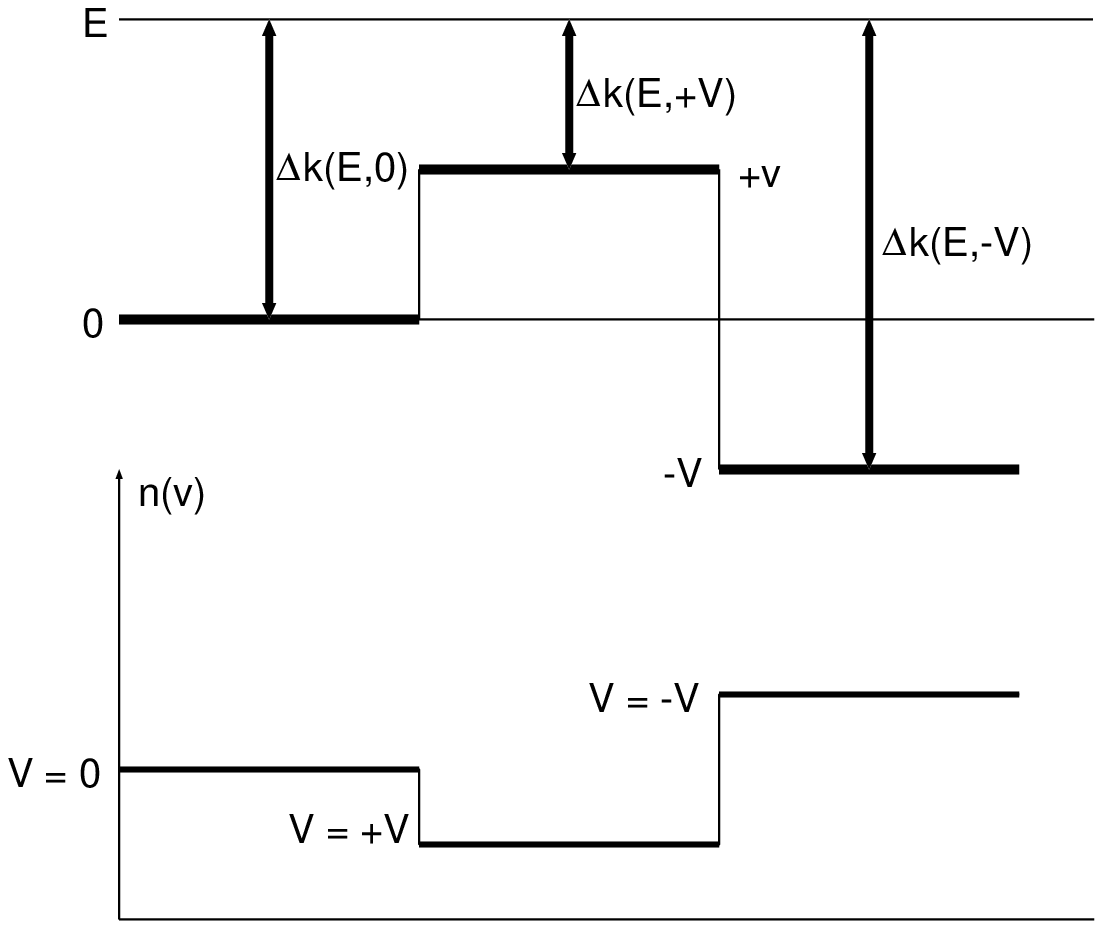} \\
\hline
\epsfxsize=0.9\hsize
\epsfbox{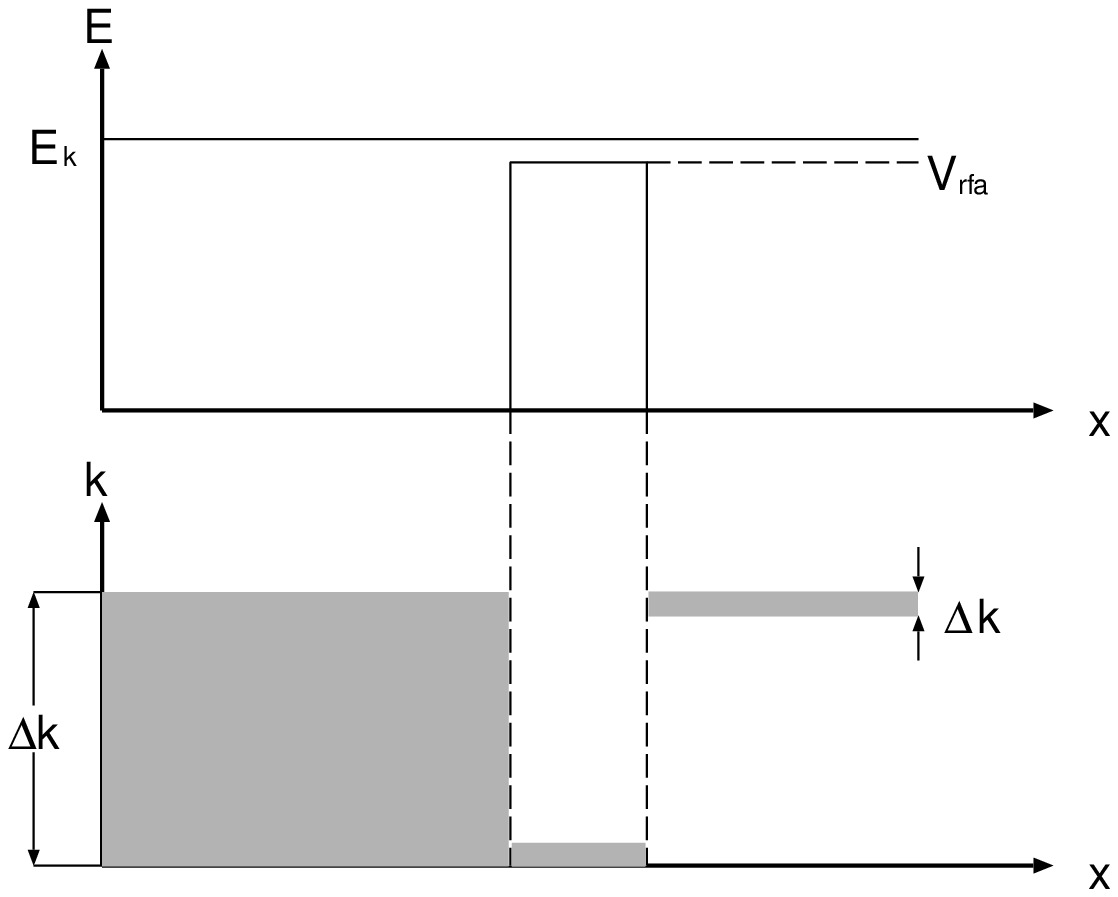} \\
\hline
\end{tabular}
\end{center}
\caption{Quantum ensembles and potentials. A potential at a specific
point $\vec{r}$ of the system changes the structure of the ensemble in
QT: a negative potential enhances, a positive diminishes the range of
allowed $k$-values (top). An energy measurement by a positive potential
(a retarding field analyzer) $V_{rfa}$
leads to a collapse of the wavefunction in $k$-space (bottom).}
\label{qe_pot}
\end{figure}

This can be done in two steps: (i) The physical environment 
determines, by way of the potential $V(\vec{r})$ and the local
boundary conditions the structure of the ensemble, i.e. the
range of intrinsic properties in a given environment.
A potential $V(\vec{r})$, for example, changes the structure of the
quantum ensemble, since it affects the range of allowed $k$-values.
For a negative potential, the range is enhanced, for a positive
one, diminished (see Fig. \ref{qe_pot}).
(ii) Once the range of intrinsic properties is determined, the
problem for a single member of the ensemble can be treated,
at this level we are dealing with a classic physical problem 
where the interactions and boundary conditions can be 
included by field theory, e.g. the wave picture of classical
electrodynamics.

This analysis of the ensemble structure in QT explains also, in a
quite natural way, how the statistical picture of QT is related to
electrodynamics: the ensembles in classical electrodynamics are
in fact quantum ensembles, where the allowed energy range is
vanishing, the energy of ensemble members is thus exactly
determined, although neither their phase nor their exact location.
This is one half of the notorious wave/particle problem, which
hounded physics since the establishment of QT (see the collection
in \cite{rauch92}): particles, in QT, are an ensemble of wave-like
structures of finite volume $V$ and a defined range of energies.
It also provides a reason for the validity of von Neumann's proof, that
QT cannot contain a theory of {\em hidden variables}, although in
quite a different sense than expected by von Neumann 
\cite{vonneumann32}: not, because quantum theory is complete, but
because quantum theory contains a - in von Neumann's words -
{\em normal} ensemble, an ensemble which cannot be described
as a sum of members of exactly determined properties (e.g.
exact location {\em and} exact energy), since the range of
allowed energy values pertains to every single
point of our system \cite{hofer97d}.

If we consider a measurement of energy on the quantum ensemble
of free electrons, e.g. by a positive potential like in 
low energy electron diffraction (LEED) experiments, it is
immediately clear that the ensemble after the potential,
assumed rectangular for simplicity, is diminished compared to
the ensemble before it. The wavefunction $\psi(\vec{r})$ has
{\em collapsed} in $k$-space (see Fig. \ref{qe_pot}).
This process, which cannot be consistently described in
the conventional formulation of QT, has led to a host of
proposed modifications, among the more daring the
{\em many world} interpretation of Everett, where every
result of a measurement occurs in a different universe
\cite{everett57}: since its publication a continuous source of inspiration
for quite a few science fiction authors. The main point here is,
that if the wavefunction is interpreted as the wavefunction of
one single particle, it must remain a mystery, how - to put it
a little sloppily - most of the particle can vanish in the
measurement, although the potential, seemingly, is not affected.
The effect is only understandable, if the ensembles underneath
a specific $\psi$ are considered.

A similar consideration applies to the notorious
{\em interaction free measurements}, where the wavefunction
of a system changes, even if no interaction occurs 
\cite{renninger60,kwiat95}. A paradoxical
consequence of this type of measurement would be, that
the energy of a system could change, even if that system
does not experience any interaction \cite{dicke81}. Within
the present theory this behavior is completely understandable,
although it points to a statistical, rather than a physical
effect: since no interaction with a particle means, in the
region where the particle is appreciable, that the wavefunction
must necessarily vanish, it excludes the existence of single
members of the ensemble in that region. Compared to the case,
where no measurement has been performed, the knowledge about
the ensemble has been changed. And since energy 
in QT is computed via the wavefunction, thus the
ensemble, the energy in the latter case can be different.
Without any spooky physical events, also without
assuming, that the {\em apparent lack of interaction ...
is only illusionary} \cite{dicke81}.

As a last example we mention the {\em quantum eraser}
measurements, where the existence of interference patterns
between two orthogonally polarized photons in a double-slit
system depends on the insertion of a polarizer with a diagonal
plane of polarization \cite{scully82}. In the conventional
framework, this behavior is attributed to the path information,
which, after the diagonal polarizer, is said to have been
''erased''. In the new framework, an identical result can been
computed by estimating the effect of polarizers on the orientation
of intrinsic field components \cite{hofer97d}.

Currently the main focus in developments is on interference
measurements, since it has been found, that a local {\em and}
causal description of this type of measurement can neither be
given in QT, nor in classical electrodynamics \cite{hofer97d}.
The main problem in electrodynamics is the result, that if
the extension of wave structures is limited, the scattering
amplitude, in a Kirchhoff approximation, contains the final 
result of measurements already at the moment, when the structure
passes the slit environment. It seems therefore, that the
mathematical formulation by way of Green's functions and
the scalar theory should be more or less algorithmic,
while actual physical processes - the interactions with the
atoms of the slit environment - are not described. Which
suggests a new theory of interferences including these 
interactions. 

\section{A sideview of Special Relativity (STR)}

It will have been noted that the total energy of an electron,
equal to $m_{e}u^2$, 
bears a slight resemblance to Einstein's energy expression
\cite{einstein05},
a resemblance, which becomes especially obvious, if photons
are considered in this model, and which possess a total energy
of $m_{ph} c^2$. It can be shown that these expressions are
more than mere coincidences, they lead, in fact, to one of the
most interesting consequences of the theory, touching a problem
known for more than fifty years and inciting the late Dirac to
qualify QED, in its present form, as a {\em very wrong theory}
\cite{dirac84}.

Since the model starts from a non-relativistic frame of reference,
a Lorentz transformation of the fundamental equations into a 
moving reference frame changes the {\em physical} state of the
system, because in this case the intrinsic potentials increase
with the electron velocity \cite{hofer97a}. In view of 
consistency, this result seemed, initially, questionable, since
it is incompatible with the relativity principle. As further
research revealed, this behavior is closely related to the
process of interaction in electrostatic fields. 

If electrostatic interactions are accomplished by photons, the 
absorption of a photon by an electron depends on proper time in
the electron system, and the acceleration is then a function of
the electron's velocity. This effect has been known for some
time. Adler remarks on that subject that {\em the time kept 
by the rapidly moving particle is dilated and hence, as the
particle's speed increases, apparently greater intervals are
taken to produce the same effect, hence the apparent increase
in resistance} \cite{adler87}. But if this is the case, then
the energy of the electron, in the limit of $n \rightarrow \infty$
absorptions

\begin{equation}
E_{n}  =  \hbar \omega_{0} \left[1 + \sum\limits_{i=0}^{n-1}
\sqrt{1 - \left( \frac{u_{i}}{c} \right)^{2}} \right] = m \, u_{n}^2
\end{equation}

will not be infinite, but converges to a finite limit,
where the total energy of the electron is, incidentally, equal
to Einstein's {\em rest energy} term $m_{e}c^2$. 

\begin{figure}
\begin{center}
\begin{tabular}{|c|}
\hline
\epsfxsize=0.9\hsize
\epsfbox{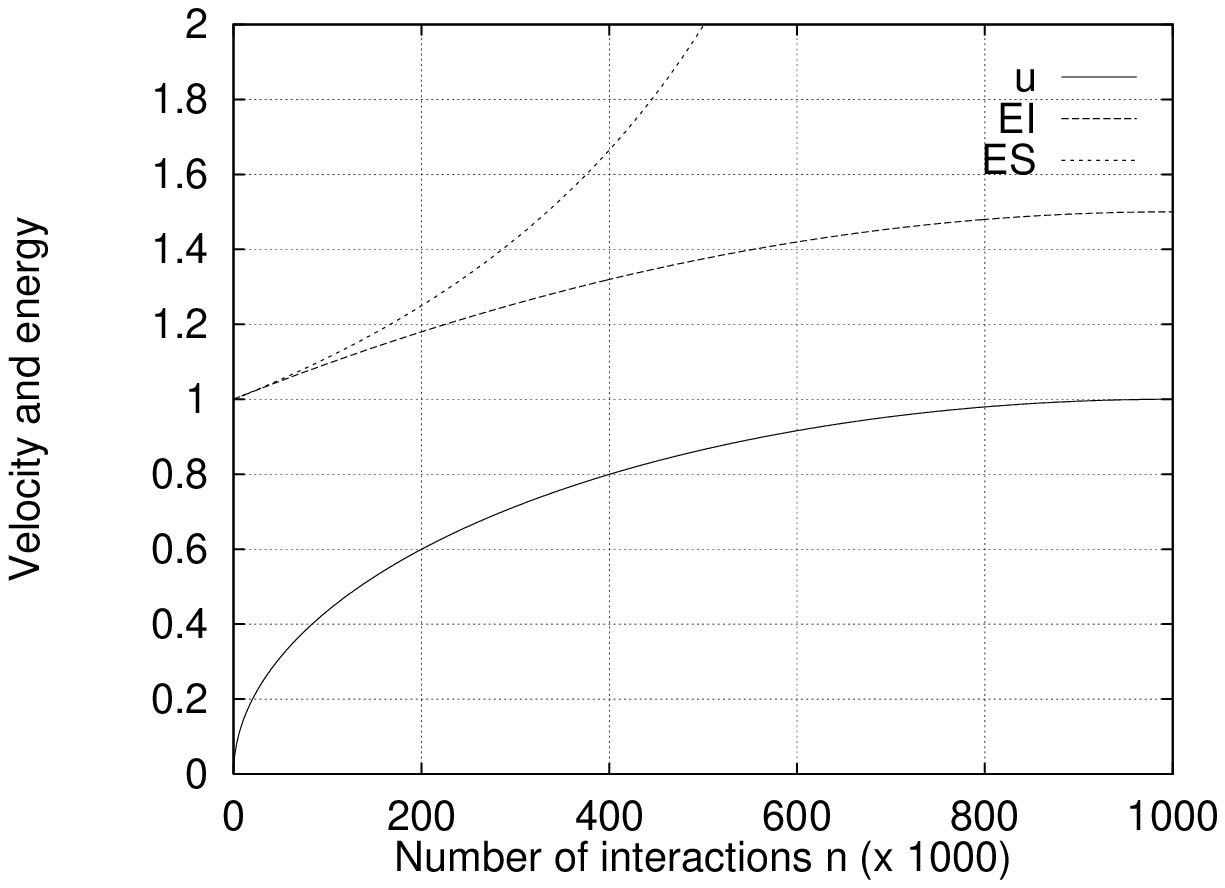}
\\ \hline
\epsfxsize=0.9\hsize
\epsfbox{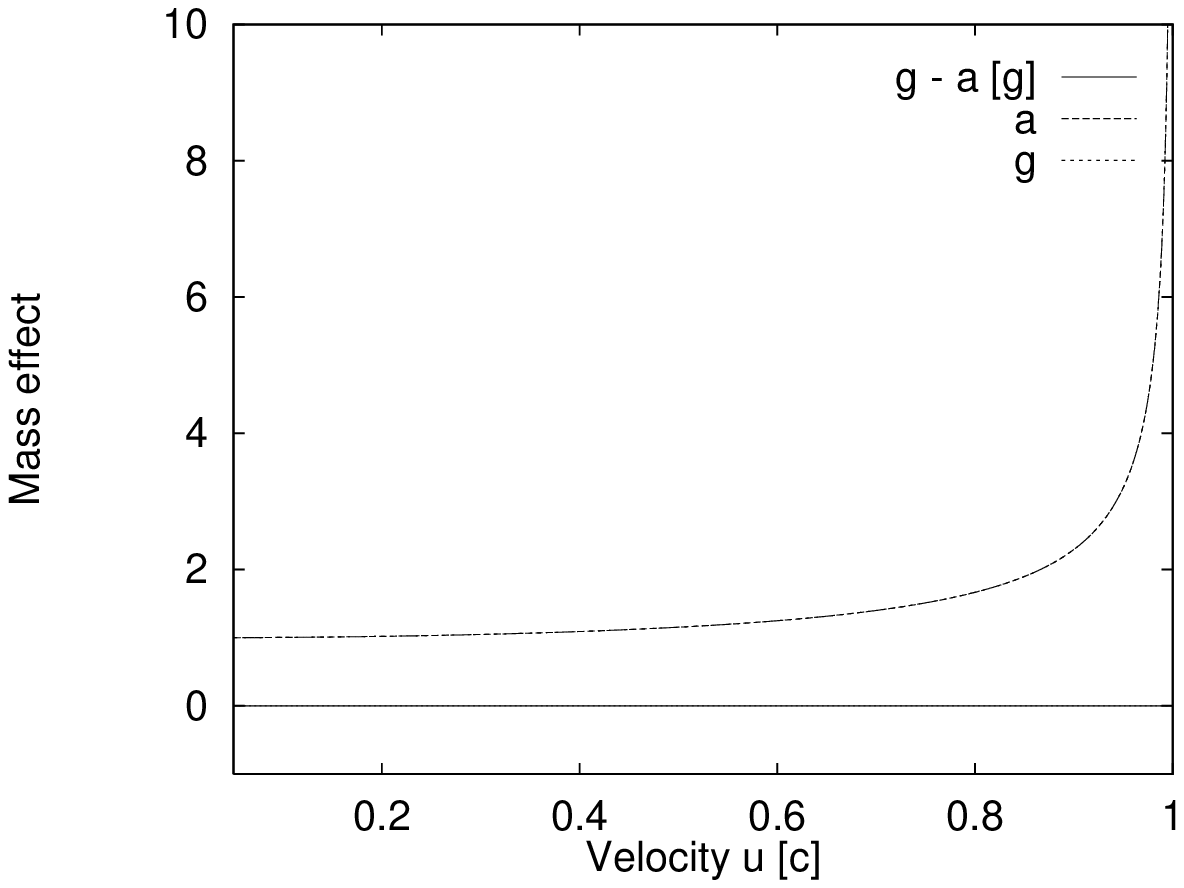}
\\
\hline
\end{tabular}
\end{center}
\vspace{0.5cm}
\caption{Electron energy due to photon absorption. Top: while in
Special
Relativity (ES) the energy of the electron becomes infinite in the limit
$u \rightarrow c$, it remains finite in the photon interaction model
(EI). Bottom: due to changed frequencies in the electron frame
accelerations decrease with increasing electron velocity. The decrease
leads to an observed but virtual increase of inertial mass. The
difference between Einstein's $\gamma$ (g) and $\alpha$ (a) is
insignificant over the whole velocity range from $u = 0.05$ to
$u = 0.99$.}
\label{fig_str}
\end{figure}

By comparing the classical
interactions due to electrostatic fields with the interactions
pertaining to photon absorption with dilated proper time it
can be established that the electron mass seems enhanced, and
that this enhancement is equal to the mass effect in STR
\cite{hofer98c}. To prove the equivalence we have
calculated the (virtual) mass enhancement due to time dilation,
described by a variable $\alpha(u)$

\begin{equation}
m (u) = \alpha m_{e} \qquad
\alpha = \sqrt{\frac{\hbar \omega_{0}}{m}} \,
\frac{\sqrt{n} - \sqrt{n-1}}{u_{n} - u_{n-1}}
\frac{u_{n}^c}{u_{n}^r}
\end{equation}

where $u_{n}^c$ and $u_{n}^r$ denote the classical and the 
interaction model velocities of the electron in an electrostatic field,
and compared $\alpha$ to Einstein's $\gamma$. The results of these
(numerical) calculations are displayed in Fig. \ref{fig_str}.
It should be noted that for reasons of comparison we have taken 
only the kinetic energy of the electron and added the rest energy.
As can be seen, the mass effect in STR coincides
nicely with the virtual mass enhancement due to time dilation.

From a physical point of view, the result means that the relativistic
mass formulas, in STR {\em artifacts of the kinematical transformation
of space and time} \cite{adler87}, are an expression of changed
interaction characteristics, described by the time dilation
in moving frames. That the result is consistent with measurements
has been shown elsewhere \cite{hofer98c}, in addition, it sheds
a new light on the so called {\em renormalization procedures}
in QED, which were the reason for Dirac's uneasiness.
As Weisskopf showed in his treatment of the free electron,
the infinite contributions to the self energy of the electron have
two origins: (i) the electrostatic energy, diverging with the radius
$a$ of the electron, and (ii) the energy due to vacuum fluctuations
of the electromagnetic fields. For these two energies $W_{st}$ and
$W_{fluct}$ he found \cite{weisskopf39}:

\begin{equation}
W_{st} = \lim_{a \rightarrow 0} \, \frac{e^2}{a} \qquad
W_{fluct} = \lim_{a \rightarrow 0} \,
\frac{e^2 h}{\pi m c a^2}
\end{equation}

The electrostatic contribution vanishes, if an electron in 
constant motion is considered, since in this case no emission
or absorption of photons will occur. The second contribution,
the vacuum fluctuations, sums up the energies 
due to the interaction of the electron with its own
created and annihilated photons in a statistical picture
which considers all possible events.

In the first calculation to master the infinity problems of
quantum electrodynamics Bethe derived the following expression
for the Lamb shift of the hydrogen electron in an
s-state \cite{bethe47}:

\begin{equation}
W_{ns}' = C \cdot \ln \frac{K}{\langle E_{n} - E_{m} \rangle_{AV}}
\end{equation}

where C is a constant
$ \langle E_{n} - E_{m} \rangle_{AV}$ the average
energy difference between states $m$ and $n$,
and K determined by the cutoff of
electromagnetic field energy. The prime refers to mass
renormalization, since the - infinite - contribution to
the electron energy due to electrostatic mass has already been
subtracted. The second infinity, the infinity of vacuum
fluctuations, is discarded by defining the cutoff K, which
in Bethe's calculation is equal to $m c^2$.
But while the energy of
the field could have any value, if the actual energy of
the electron has a singularity at $u = c$
(and K could therefore be infinite), this is not the
case if the energy remains finite in this limit: in this case
the total energy difference between a relativistic electron and
an electron at rest is $m c^2$ according
to our calculations. This is, incidentally,
equal to the rest energy of the electron. It seems,
therefore, that the renormalization procedures
\cite{schweber94} may have their ultimate justification
in finite electron energy.

It should be noted that 
it is not yet sufficiently clear, from the viewpoint of this
new theory, how the more subtle theoretical developments of
QED shall be put into the new framework. In addition, it has
been seen by reanalyzing experiments and their description in
the standard theory, that progress in not to be expected by an
{\em equation for everything}. Rather by careful revision of
experimental evidence and subtle speculation within the new
framework: a tedious task, it seems, but which is the
price paid for the insight gained into fundamental processes.

\section{Conclusions}

We have shown in this paper that a new theoretical framework, based
on the intrinsic properties of electron, is suitable to remove the
notorious infinity problems in QED and to describe a realistic
electron in accordance with experimental data.
Electrostatic and magnetic interactions have been treated in this
framework, and the origin of the {\em non-local} properties of
particle spin has been determined. We have also described the
borderline between the usually statistical interpretation of
the wavefunction and the, physically relevant, intrinsic wave
properties. In this case a novel structure of the ensembles in
quantum theory was proposed, which is due to the omission of
intrinsic enery components. Finally, we have described how the
theory treats photon absorption processes in a relativistic
context, which led to the conclusion that the mass enhancement
in the electron system is only virtual and an effect of time
dilation. It was shown, how this result lies underneath the
hitherto unexplained renormalization procedures in relativistic
quantum field theory.

\section{Acknowledgements}

I'd like to thank Prof. Dvoeglazov for his kind invitation
to contribute to this volume.




\begin{references}
\bibitem{schweber94}
S. S. Schweber, {\em QED and the Men Who Made It},
Princeton (1994)
\bibitem{schweber61}
S. S. Schweber {\em An Introduction to Relativistic Quantum Field
Theory}, Evanston (1961)
\bibitem{barut97}
A. O. Barut, in D. Hestenes and A. Weingartshofer (eds.)
{\em The Electron: New Theory and Experiment}, Kluwer, Dordrecht
(1991) 
\bibitem{bunge55}
M. Bunge, {\em Nuovo Cimento} {\bf 1}, 977 (1955)
\bibitem{recami97}
E. Recami and G. Salesi,
'' Kinematics and hydrodynamics of spinning particles:
Some simple considerations'', in
J. Keller and Z. Oziewicz (eds),
{\em The Theory of the Electron}, UNAM (1997)
\bibitem{bohm52}
D. Bohm, {\em Phys. Rev.} {\bf 85}, 166 (1952); 180 (1952)
\bibitem{hofer95}
W. A. Hofer {\em Spec. Sci. Tech.} {\bf 18}, 157 (1995)
\bibitem{hofer97a}
W. A. Hofer 
''Internal Structures of Electrons and Photons and
some Consequences in Relativistic Physics'', F. Selleri
(ed) {\em Open Questions in Relativistic Physics},
Montreal (1998)
\bibitem{hofer97b}
W. A. Hofer {\em Spec. Sci. Tech} {\bf 20}, 115 (1997)
\bibitem{hofer97c}
W. A. Hofer {\em Non--locality of particle spin:
a consequence of interaction energy?} 
 preprint at quant-ph/9702023
\bibitem{hofer97d}
W. A. Hofer {\em Measurement processes in quantum physics: a new
theory of measurements in terms of statistical ensembles}
 preprint at quant-ph/9704006
\bibitem{hofer98a}
W. A. Hofer {\em A dynamic model of atoms: structure,
internal interactions and photon emissions of hydrogen}
 preprint at quant-ph/9801044
\bibitem{hofer98b}
W. A. Hofer {\em Physica A} {\bf 256}, 178 (1998)
\bibitem{hofer98c}
W. A. Hofer {\em Electron acceleration due to photon absorption:
A possible origin of the infinity problems in relativistic 
quantum fields} preprint at quant-ph/9805061
\bibitem{mcgregor92}
M. H. MacGregor {\em The Enigmatic Electron.}, Kluwer,
Dordrecht (1992)
\bibitem{schrodinger26}
E. Schr\"odinger, {\em Ann. Physik} {\bf 79}, 361; 489 (1926)
\bibitem{heisenberg27}
W. Heisenberg, {\em Z. Physik} {\bf 43}, 172 (1927)
\bibitem{debroglie82}
L. de Broglie, \mbox{\it Les incertitudes} {\it d'Heisenberg et l'interpretation
probabiliste de la mechanique ondulatoire}, Gauthier-Villars, Paris
(1988)
\bibitem{jackson84}
J. Jackson, {\em Classical Electrodynamics}, Wiley \& Sons,
New York (1984)
\bibitem{aspect82}
A. Aspect, {\em Phys. Rev. Lett} {\bf 49}, 91; 1804 (1982)
\bibitem{claverie80}
P. Claverie and S. Diner, {\em Israeli J. Chem.}
{\bf 19}, 54 (1980)
\bibitem{kracklauer98}
A. F. Kracklauer, preprint at QUANT-PH 9804059
\bibitem{rauch92}
H. Rauch, ''Neutron interferometric tests of quantum mechanics'',
in F. Selleri (ed) {\em Wave-particle duality}, Plenum Press,
New York (1992)
\bibitem{zeilinger81}
A. Zeilinger, R. Gaehler, C. G. Shull, and W. Teimer,
{\em Symposion on Neutron Scattering, Argonne National Laboratory},
Am. Inst. Phys. (1981)
\bibitem{hofer96}
W. A. Hofer {\em Beyond Uncertainty: the internal structure of electrons
and photons}, preprint at quant-ph/9611009
\bibitem{bender84}
D. Bender et. al. {\em Phys. Rev D} {\bf 30}, 515 (1984)
\bibitem{bohm66}
D. Bohm and J. Bub 
{\em Rev. Mod. Phys.} {\bf 38}, 453 (1966)
\bibitem{born26}
M. Born, {\em Z. Physik} {\bf 37}, 863 (1926)
\bibitem{cramer86}
J. G. Cramer {\em Rev. Mod. Phys.} {\bf 58}, 647 (1986)
\bibitem{vonneumann32}
J. von Neumann, {\em Mathematische Grundlagen der Quantenmechanik},
Springer, Berlin (1932)
\bibitem{everett57}
H. Everett, {\em Rev. Mod. Phys.} {\bf 29}, 454 (1957)
\bibitem{renninger60}
M. Renninger, {\em Z. Physik} {\bf 158}, 417 (1960)
\bibitem{kwiat95}
P. Kwiat, H. Weinfurter, T. Herzog, A. Zeilinger and
M. A. Kasevich, {\em Phys. Rev. Lett.} {\bf 74}, 4763 (1995)
\bibitem{scully82}
M. O. Scully, and K. Dr\"uhl, {\em Phys. Rev. A} {\bf 25},
2208 (1982)
\bibitem{dicke81}
R. H. Dicke, {\em Am. J. Phys.} {\bf 49}, 925 (1981)
\bibitem{einstein05}
A. Einstein, {\em Ann. Physik} {\bf 18}, 639 (1905)
\bibitem{dirac84}
P. A. M. Dirac, {Eur. J. Phys.} {\bf 5}, 65 (1984)
\bibitem{adler87}
C. G. Adler, {\em Am. J. Phys.} {\bf 55}, 739 (1987)
\bibitem{weisskopf39}
V. F. Weisskopf  {\it Phys. Rev.}\, {\bf 56}, 72 (1939)
\bibitem{bethe47}
H. A. Bethe  {\it Phys. Rev. } \, {\bf 72}, 339 (1947)
\end{references}
\end{document}